\documentclass[twocolumn,preprintnumbers,amsmath,amssymb]{revtex4}

\usepackage{graphicx}  
\usepackage{sidecap}
\begin{document}
\title{Amplification
 of  surface plasmons  in graphene-black phosphorus  injection laser heterostructures 
}
\author{V. Ryzhii$^{1,2,3,4}$,   T.~Otsuji$^1$,  M. Ryzhii$^5$,     A. A. Dubinov$^6$,    V. Ya. Aleshkin$^6$,  V. E. Karasik$^4$,
and M. S. Shur$^7$}
\address{
$^1$Research Institute of Electrical Communication, Tohoku University, Sendai 980-8577, Japan\\
$^2$Institute of Ultra High Frequency Semiconductor Electronics of RAS,\\
 Moscow 117105, Russia\\
 $^3$  Center for
Photonics and Two-Dimensional  Materials, Moscow Institute of Physics Technology,
Dolgoprudny 141700, Russia\\
 $^4$Center for Photonics and Infrared Technology, Bauman Moscow State Technical University, Moscow 111005, Russia\\
 $^5$Department of Computer Science and Engineering, University of Aizu, Aizu-Wakamatsu 965-8580, Japan\\
 $^6$Institute for Physics of Microstructures of RAS and Lobachevsky University of Nizhny Novgorod,
Nizhny Novgorod, 60395, Russia\\
$^7$ Department of Electrical, Computer, and Systems Engineering, Rensselaer Polytechnic Institute, Troy, New York 12180, USA\\
}
 \begin{abstract} 
We propose and evaluate the heterostructure based on the graphene-layer (GL) with the lateral electron injection 
from the side contacts and
the hole vertical injection via the black phosphorus layer (PL) (p$^+$PL-PL-GL heterostructure). 
Due to a relatively small energy of the holes injected  from the PL into the GL (about 100 meV,  smaller than the energy of optical phonons in the GL which is about 200 meV),
the hole injection can effectively cool down
 the two-dimensional electron-hole plasma in the GL.  This simplifies the realization of the interband population inversion and the achievement
of the negative dynamic conductivity in the terahertz (THz) frequency range enabling the amplification of the surface plasmon modes. The later can lead to the plasmon lasing. The conversion of the plasmons into the output radiation  can be used for a new types of the THz sources.
\end{abstract} 

\maketitle

\newpage

\section{Introduction}

The gapless energy spectrum of graphene layers (GLs)~\cite{1,2} enables their use in the interband photodetectors~\cite{3,4,5,6} (see also the review articles~\cite{7,8,9,10,11} and the references therein) and sources (for example,
~\cite{11,12,13,14,15,16,17,19,20,21,22,23,24,25,26,27,28} operating in  the terahertz (THz) a far-infrared (FIR) spectral ranges.
In particular, the optical~\cite{12,14,17,18,19,20,21}   and lateral injection pumping of the GLs from the side 
n- and p-contacts (i.e., from the chemically- or electrically-doped regions)~\cite{13,16,22,24,25,26,27} 
 can lead to the interband population inversion and negative dynamic conductivity.
 This can enable the THz lasing experimentally demonstrated.
 The GL-based heterostructure with lateral current injection and the grating providing the distributed feedback exhibits 
  a single-mode lasing at 5.2 THz and a broadband (1 - 8 THz) amplified spontaneous emission both at 100~K ~\cite{24,25,26}.
 To increase the operating temperature and   further enhance  the THz gain and lasing radiation intensity,
  the injection efficiency should be elevated.

The advantage of the carrier lateral double injection pumping from the side n- and p-contact regions  in the GL-structures~\cite{13,16},
in comparison with the optical pumping is associated with relatively low energies of the injected carriers.
  While the energy of the injected carriers is about $\varepsilon_i \simeq T_0$~\cite{30,31}, the initial energy 
of the photogenerated carries is equal to $\varepsilon_{Opt} = \hbar\Omega/2$~\cite{12,28,32}. Here $T_0$ is the lattice temperature, $\hbar\Omega$ is the energy of photons in the incident (pumping) radiation. In practical devices with the optical pumping using  A$_3$B$_5$ semiconductor interband lasers integrated with the GL-structure, $\hbar\Omega \sim 1$~eV. In the case of optical pumping
by mid-IR quantum-cascade lasers, $\hbar\Omega$ can be markedly smaller, but the integration of the pumping source with the GL
can be challenging due to the radiation polarization problems. 
The relatively high values of $\varepsilon_{opt}$ determine  rather high effective temperature $T$ of the photogenerated two-dimensional 
electron-hole plasma (2DEHP) in the GL complicating the achievement of the strong interband population inversion and lasing~\cite{32}.

\begin{figure*}[t]
\centering
\includegraphics[width=14.0cm]{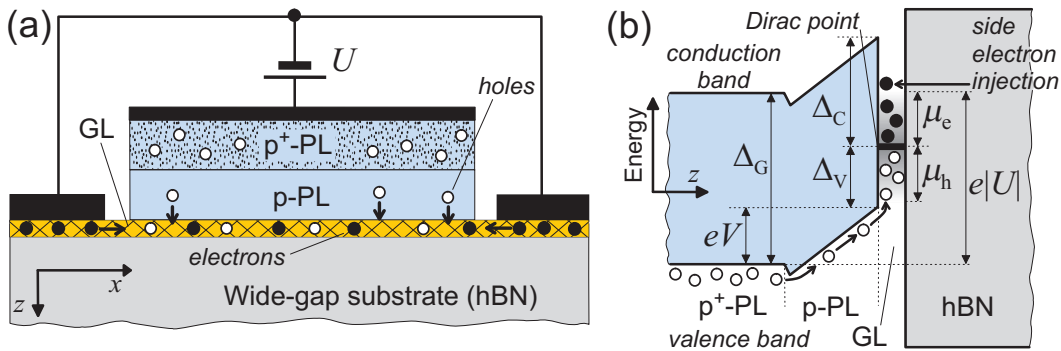}
\caption{Schematic view of (a) the p$^+$P-PL-GL heterostructure and (b) its band diagram at a voltage $U$. }
\label{F1}
\end{figure*}

The efficiency of the  lateral injection can be impaired by a decrease in the carrier density in the GL-heterostructure center caused by recombination (the sag of the carrier spatial lateral distribution~\cite{30}, which weakens the population inversion and decreases the net  THz gain .
This limits the lateral size of the device (spacing between the side n$^+$- and p$^+$-contacts to the GL
by the carrier lateral ambipolar diffusion length. 
Shortening of the active part of the GL increases the leakage currents (electrons 
and holes reaching  the p-contact and n-contact, respectively). 

A compromise can be reached  using of the lateral injection of one type of carriers (the electron injection
from the side n$^+$-contacts) and the vertical injection
of the other type (the hole injection via the bulk p-layer).
A proper band alignment of the GL and the bulk material layer serving as the vertical injector could
 minimize  or even avoid the 2DEHP heating by the injection of hot holes.
This implies that  the material for the hole injector should  have the energy spacing,
$\Delta_V$,  between the Dirac point
in the GL and the valence band top of the injector material   as small as possible. One of such candidates for the injector  material is the black phosphorus
~\cite{33,34,35,36,37,38}. This material is now considered to be very promising for different electronic and optoelectronic devices applications (see, for example,~\cite{33,34,35,36,37,38,39,40,41,42,43,44,45,46,47,48,49,50,51}). The quantity $\Delta_V$ in the black phosporous layers (PLs) comprising several
 atomic sheets is estimated as $\Delta_V \simeq 100$~meV with the energy band $\Delta_G = \Delta_V + \Delta_C \simeq 300$~meV ($\Delta_C$ is the GL-PL electron affinity). Since the energy of the holes injected into the GL from the PL, $\sim \Delta_V$,  is smaller than the energy of optical phonons in the GL (about of 200 meV), the hole injection can even cool in a substantial cooling of the 2DEHP (in contrast to the 2DEHP heating  in the case of the injection from materials with $\Delta_V > \hbar\omega$).
The latter is definitely beneficial for the 2DEHP degeneration and, hence, for a stronger population inversion.  A high dc conductivity of the 2DEHP in the GL  provides a fairly weak sag in the carrier densities  at the  pumping method in question, so that the spacing between the side contacts can be
fairly large resulting in a decrease of the leakage currents. All  this is useful for an enhancement of the output THz power in the lasers based
on the GL-PL heterostructures with the combined pumping. 

Apart from the unique electron and hole  properties of the PLs in the in-plane directions, the PLs exhibit a rather good carrier transport in the direction perpendicular to the phosphorous atomic sheets. This makes the PLs very suitable for the hole injectors in the PL-GL lasers.
As demonstrated recently, the quantity $\Delta_V$  in the devices in question can be even smaller  if
the PL is replaced by black arsenic-phosporous compounds~\cite{52}.

In this paper, we propose and analyze the GL-based heterostructure with the lateral injection of electrons from the side 
n$^+$ contacts and the vertical injection of holes from the bulk  p$^+$PL-PL-GL -structure.     
We calculate the dependences of the carrier effective temperature, their quasi-Fermi energies, 
the 2DEHP  frequency-dependent dynamic conductivity, and the  coefficient of the surface plasmonic modes amplification
 as  functions of the injected current for different structural parameters. Using these data, we find the conditions at which this conductivity is negative, and
the coefficient of the surface plasmons amplification is positive. 
The  plasmonic modes  self-excitation in the latter case can lead to the plasmonic lasing followed by
the conversion of these modes into the output THz radiation.

The cooling of the 2DEHP under the vertical 
injection might lead  a substantial softening of the population inversion conditions and the conditions of the amplification and self-excitation
of the photonic and plasmonic modes. Therefore,  
the proposed heterostructure can serve as an active part of the THz and FIR lasers with the photonic and plasmonic wave guides.

\section{Device structure}

Figure~1 shows the schematic view of
this heterostructure with a relatively narrow-gap injector p-type black PL, GL, on  a wide-gap substrate  and its  energy diagram at the operating bias voltage $U$ ($|U| > V_{bi} \simeq \Delta_V/e$, where $V_{bi}$ the built-in voltage). 
As for the substrate, several relatively wide-gap materials can be used, in particular, hexagonal Boron Nitride (hBN) because the GLs on the hBN substrate exhibit exceptionally high  mobility values.
 A wide gap in the hBN substrate
provides high energy barrier for the electrons and holes in the GL and blocks their leakage to the substrate.
At the applied bias voltage, the electron can freely fill in the GL conduction band, while the  holes pass vertically from the heavily-doped p$^+$ region through the undoped or lightly doped
layer and are injected into the GL. Due to the energy spacing, $\Delta_V$, between the valence band of the hole injector
and the Dirac Point in the GL, the injected holes injected  bring a substantial energy into the electron-hole system in the GL, but this energy is effectively removed due to the emission of the high-energy (about 200~meV)
 optical phonons in the GL  This can result in the cooling 
of the  2DEHP injected into the GL.
   
The device model used for the calculation accounts for a strong  deviation of the 2DEHP  from equilibrium caused the  injection.  
The efficient carrier-carrier interaction in
a high density of the 2DEHP 
  leads to the "Fermization" of the carrier energy distributions, 
  so that electrons and holes can be described by the Fermi functions with the same effective temperature
T =$T_e = T_h$ and the quasi-Fermi energies $\mu_e$ and $\mu_h$,  which might differ from  
their equilibrium values.
At temperatures close to the room temperature, the carrier interactions with the optical phonons in the GL can are   the main mechanism of the energy relaxation and recombination~\cite{32,53}. The surface optical phonons at the GL-hBN interface can play a significant role in the relaxation of nonequilibrium carriers in the GL~\cite{54}. 
The direct Auger processes in the GLs are virtually prohibited~\cite{55} due to the linearity of the carrier energy spectra~\cite{1}. More complex Auger processes are  also  effectively suppressed~\cite{56}
The role of the Auger interband processes will be briefly  considered in the Appendix.

\section{Energy and density balances in  the 2DEHP}

In each act of the interband and intraband  emission/absorption  of the GL optical phonons 
(with the energy $\hbar\omega_0 \simeq 200$~meV
and the interface optical phonons (with the energy $\hbar\omega_S \simeq 100$~meV)
 the energy of the 2DEHP
decreases/increases by the quantity  $ \hbar\omega_0$.  
The resulting the energy balance equation is

\begin{eqnarray}\label{eq1}
\exp\biggl(\frac{\mu_e + \mu_h}{T} \biggl)
\exp\biggl[\hbar\omega_0\biggl(\frac{1}{T_0} - \frac{1}{T}\biggr) \biggr] -1\nonumber\\
+ s\frac{\omega_S}{\omega_0}\biggl\{\exp\biggl(\frac{\mu_e + \mu_h}{T} \biggl) \exp\biggl[\hbar\omega_S\biggl(\frac{1}{T_0} - \frac{1}{T}\biggr)\biggr] - 1\biggr\}\nonumber\\
+a\biggl\{\exp\biggl[\hbar\omega_0\biggl(\frac{1}{T_0} - \frac{1}{T}\biggr)\biggr] -1\biggr\}\nonumber\\
 + a s\frac{\omega_S}{\omega_0}\biggl\{\exp\biggl[\hbar\omega_S\biggl(\frac{1}{T_0} - \frac{1}{T}\biggr)\biggr]    - 1\biggr\}
 = 
 \frac{j}{j_G}\biggl(\frac{\Delta_i}{\hbar\omega_0} \biggr).
\end{eqnarray}
The equation governing the electron and hole  balance  is given by:
\begin{eqnarray}\label{eq2}
\exp\biggl(\frac{\mu_e + \mu_h}{T} \biggl)
\exp\biggl[\hbar\omega_0\biggl(\frac{1}{T_0} - \frac{1}{T}\biggr) \biggr] - 1\nonumber\\
+ s \biggl\{\exp\biggl(\frac{\mu_e + \mu_h}{T} \biggl)\exp\biggl[\hbar\omega_S\biggl(\frac{1}{T_0} - \frac{1}{T}\biggr) \biggr] - 1\biggr\}
 = \frac{j}{j_G}.
\end{eqnarray}
Here  $j$ is the injection current density,$j_{G} = e\Sigma_0/\tau_{Opt}^{inter}$,  $\Sigma_0$ is the  characteristic carrier density determined by the energy dependence of the density of state in the GL near the Dirac point, $e$ is the electron charge,   $a =\tau_{Opt}^{inter}/\tau_{Opt}^{intra} $
. 
is the ratio of the pertinent times characterizing the interband transitions, $as = \tau_{Opt}^{inter}/\tau_S^{intra}$,
$\tau_{Opt}^{inter}$ and $\tau_{Opt}^{intra}$ are the characteristic recombination and intraband relaxation
times associated with the carrier interaction with the optical phonons
($\tau_{Opt}^{inter} < \tau_{Opt}^{intra} $~\cite{32}), $\tau_{S}^{inter}$ and $\tau_{S}^{intra}$ are the same times but associated with the surface optical phonons,
the quantity $\Sigma_0/\tau_{Opt}^{intra}$ is the order of the electron-hole pair thermogeneration rare per unit area in equilibrium, so that $\tau_{Opt}^{intra} \sim \tau_0\exp(\hbar\omega_0/T_0)$, where $\tau_0$ is the time of the optical phonon spontaneous emission,   
$T_0$ is the lattice temperature, $\Delta_i = \Delta_V + 3T_P/2$ is the average energy bringing by the hole injected from the BL to the GL  (see Appendix A), and $T_P$ is the effective hole temperature in the PL (near the PL-GL interface.
 
 The terms in the left-hand sides of Eqs.~(1) and (2) describe the processes of the interband and intraband energy relaxation and the recombination-generation processes. The right-hand side terms
correspond to the energy  and carrier fluxes into the GL associated with the injection.
 Equations~(1) and (2) are the versions of the pertinent equations used previously (for example,~\cite{16,32}) and generalized to take into account for  two types of optical phonons (the optical phonons in the GL and the surface optical phonons at the GL-hBN interface).
 
 \section{Effective temperature and quasi-Fermi energies as functions of the injected current}
 
In the limit of small $s$, which could correspond to the device with the substrate (instead of the hBN substrate) exhibiting very weak interaction of its phonon system with the carriers in the GL   from Eqs.~(1) and (2) we obtain
 
\begin{eqnarray}\label{eq3}
\frac{1}{T} = \frac{1}{T_0} \biggl\{1 - \frac{T_0}{\hbar\omega_0}\ln \biggl[1 +\frac{j}{j_G}\biggl(\frac{\Delta_i}{\hbar\omega_0} - 1\biggr)\frac{1}{a}\biggr] \biggr\},
\end{eqnarray}

\begin{eqnarray}\label{eq4}
\frac{\mu_e + \mu_h}{T} = 
\ln\biggl[\frac{1 +  \displaystyle\displaystyle\displaystyle\frac{j}{j_G}}
{1  +\displaystyle\frac{j}{j_G}\biggl(\frac{\Delta_i}{\hbar\omega_0} - 1\biggr)\frac{1}{a}}\biggr].
\end{eqnarray}

From Eq.~(3)  one can see that
$T \geq T_0$ if $\Delta_i > \hbar\omega_0 \simeq 200$~meV (heating of  the 2DEHP  by the injection current) and $T < T_0$ (cooling of this plasma) if $\Delta_i < \hbar\omega_0$.
Simultaneously, from Eq.~(4) we find that $\mu_e + \mu_h < 0$ and  $\mu_e + \mu_h > 0$
when $\Delta_i/\hbar\omega_0 > 1+a$ and  $\Delta_i/\hbar\omega_0 < 1+a$, respectively.
In the case $1 < \Delta_i/\hbar\omega_i < 1 + a$, both $(T-T_0)$
and $(\mu_e + \mu_h)$ are positive.

If $\Delta_i > \hbar\omega_0$, an increase in the injected current density $j$ results in a monotonic rise of the effective temperature. In this case, Eq.~(3) yields the $T - j$ dependence, which diverges at a fairly large value $j = j_{\infty}$, where 

\begin{equation}\label{eq5}
j_{\infty} = j_G \frac{a[\exp(\hbar\omega_0/T_0 - 1)]}{(\Delta_i/\hbar\omega_0) - 1} \simeq  j_G \frac{a\exp(\hbar\omega_0/T_0)}{(\Delta_i/\hbar\omega_0) - 1}.
\end{equation}
Such a divergence  means that at such a pumping the interaction of the carriers with optical phonons in the GL is not able to transfer the energy  brought to the GL  by the injected carriers to the optical phonon system.
In reality, a sharp increase in  the effective temperature might be limited
 by additional  energy relaxation mechanisms engaging  at very large temperatures.
 
 When $j$ tends to $j_{\infty}$, from Eq.~(4) we obtain  
 
 \begin{equation}\label{eq6}
\frac{\mu_e + \mu_h}{T}  \simeq \ln \biggl(\frac{a}{\Delta_i/\hbar\omega_0 - 1} \biggr).
\end{equation}
The latter quantity can be both positive   and negative ((degenerate and nondegenerate 2DEHP, respectively).

In the most interesting case   $\Delta_i  < \hbar\omega_0$,  $j$ tends to the saturation current density 

\begin{equation}\label{eq7}
j_{sat} = j_G \frac{a}{(1 - \Delta_i/\hbar\omega_0)},
\end{equation}
and the effective temperature $T$ steeply drops tending to zero.
 Apart from this, at $j \simeq  j_{sat}$,  the ratio $(\mu_e + \mu_h)/T$  tends to infinity, while $(\mu_e + \mu_h)$
 tends to $\hbar\omega_0$.  In such a case, the hole quasi-Fermi energy can become close to $\Delta_V$. The latter,  accompanied with a strong decrease in the effective temperature (and, hence, a strong carrier system degeneration),  leads to a dramatic suppression of the hole capture
 into the GL because the GL valence band becomes overfilled up to the top of the barrier ($\mu_h \simeq \hbar\omega_0/2 \sim \Delta_V$). 
   As a result, the injected current density can not markedly exceed $j_{sat}$ (the injected current saturation).

At $T = 300$K, setting~\cite{53} $\Sigma_0/\tau_{Opt}^{inter} \simeq 10^{21}$~cm$^{-2}$s$^{-1}$. 
we obtain
$j_G= e\Sigma_0/\tau_{Opt}^{inter} = 1.6\times 10^{2}$~A/cm$^2$. The quantity $j_0$ can be of the same order of magnitude as $j_G$.

Equation~(2) yields the sum of the electron and hole quasi-Fermi energies $\mu_e + \mu_h$ versus the injected (recombination)
current $j$. An additional  relationship between $\mu_e$ and  $\mu_h$ on the  one hand and $j$ on the other
can be obtained considering the difference in the electron and hole densities, $\Sigma_e$ and $\Sigma_h$, in the GL
determined by the electric field $E_{PG}$ at the PL and GL interface.  Using Eq.~(A6), we obtain

\begin{equation}\label{eq8}
 \Sigma_e - \Sigma_h = \frac{\kappa\,V}{4\pi\,ed} = \frac{\kappa}{4\pi\,e^2b_PN_a}\,j.
\end{equation}
where $\kappa = (\varepsilon_P + \varepsilon_{hBN})/2$ is the effective dielectric constant determined by the  dielectric constants of the layers ($\varepsilon_P$ and $\varepsilon_{hBN}$ are the dielectric constants of the BL and hBN,respectively)  sandwiching the GL and $b_P$ is the hole mobility in the direction perpendicular to the heterostructure plane. Considering that the electron and hole densities in the GL are related to the quasi-Fermi energies (of the degenerate electron and hole components, $\mu_e, \mu_h > T$) as 
$\Sigma_e \simeq \mu_e^2/\pi\hbar^2v_W^2$ and  $\Sigma_h \simeq \mu_h^2/\pi\hbar^2v_W^2$,
where $v_W \simeq 10^8$~cm/s is the characteristic carrier velocity in the GLs, 
from Eq.~(8) we arrive at (see also Appendix B)

\begin{equation}\label{eq9}
 (\mu_e - \mu_h) (\mu_e + \mu_h)= \frac{\kappa\hbar^2v_W^2}{4e^2b_PN_a}j = T_0^2D\frac{j}{j_G}. 
\end{equation}
where
\begin{equation}\label{eq10}
 D = \frac{\kappa\hbar^2v_W^2\Sigma_0}{4eb_PN_a\tau_{Opt}^{inter}T_0^2} =\frac{\kappa\hbar^2v_W^2m\Sigma_0}{4e^2N_a\tau_P\tau_{Opt}^{inter}T_0^2}.
 \end{equation}

For $\kappa \simeq 6$, $b_P = (250 - 500)$~cm$^2$/V$\cdot$s ant $N_a = 5\times 10^{15}$~cm$^{-3}$, Eq.~(7) yields $D \simeq 0.019 - 0.038$. 

Figure~2 shows  the dependences of the carrier effective temperature $T$ in the GL, their net quasi-Fermi energy
$(\mu_e + \mu_h)$, and the ratio 
$(\mu_e + \mu_h)/T$
on the normalized injection current density $j/j_G$ calculated using Eqs.~(3) and (4), i.e., neglecting the contribution of the surface optical phonons ($s = 0$),  for different values $\Delta_V$. We set $\hbar\omega_0 = 200$~meV,
$T_0 = 25$~meV, and $a = 0.25$.

\begin{figure}[t]
\centering
\includegraphics[width=6.0cm]{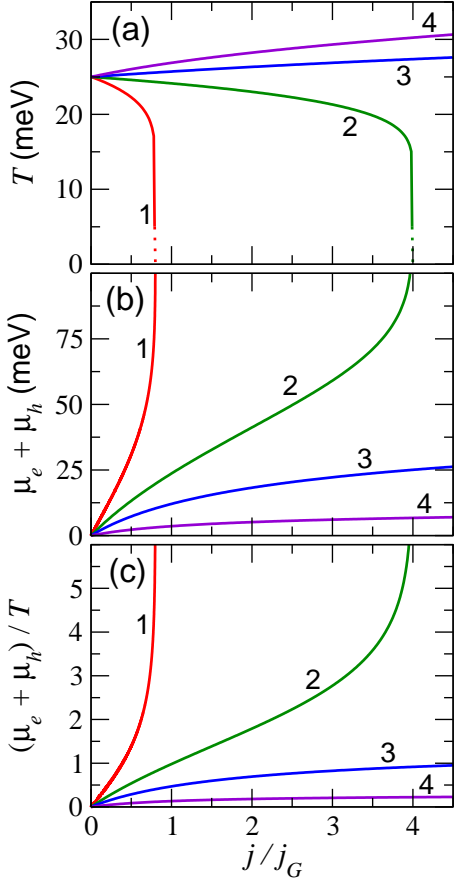}\\
\caption{The dependences of (a)  carrier effective temperature $T$, (b)  the net quasi-Fermi energy
$(\mu_e + \mu_h)$, and (c) the ratio 
$(\mu_e + \mu_h)/T$ on the normalized injection current density $j/j_G$ for different $\Delta_V$: 1 - $\Delta_V
= 100$~meV, 2 - $\Delta_V = 150$~meV, 3 - $\Delta_V = 175$~meV,1 - $\Delta_V
= 200$~meV.}
\label{F2}
\end{figure}

The plots in Figure~2 confirm the above qualitative analysis  of the effective temperature and the quasi-Fermi energies behavior as functions of the injected current density. In particular,
Fig.~2 demonstrates the possibility of a fairly strong cooling and degeneration of the 2DEHP in the
GL  with increasing injection current density providing that  $\Delta_i < \hbar\omega_0$ (curves "1" and "2"). But at $\Delta_i < \hbar\omega_0$ Fig.~2 (curves "3" and  "4" ) demonstrates a moderate 2DEHP heating, which, nevertheless,  is accompanied with the 2DEHP degeneration, although the latter is also moderate.

\begin{figure}[t]
\centering
\includegraphics[width=6.0cm]{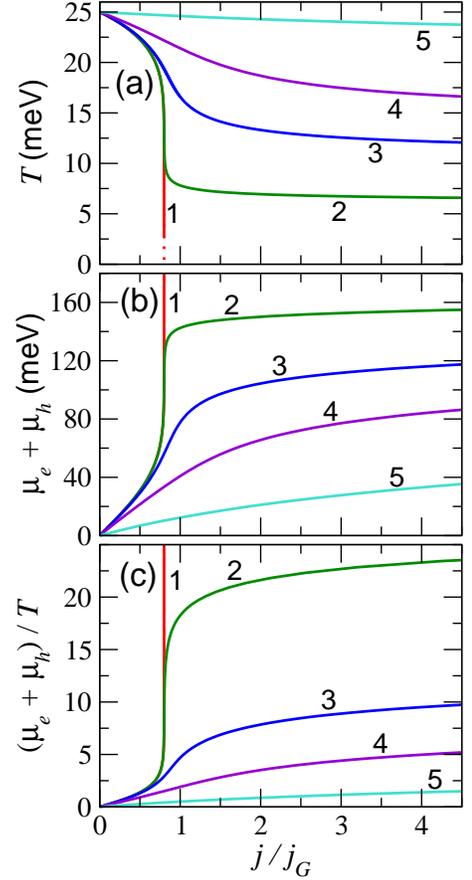}\\
\caption{The same as in Fig.~2 but  for values of the parameter $s$ characterizing the relative strength of the carrier interaction with the surface phonons: 
 $\Delta_V= 100$~meV,  1 - $s = 0$;\, 2 - $s =0.001$;\, 3- $s =0.01$;\, 4 - $s = 0.1$, and 5 - $s = 1.0$.}
\label{F3}
\end{figure}

The inclusion an extra intraband and interband relaxation mechanism, like that associated with the carrier interaction with surface optical phonons ($s \neq 0$) with $\hbar\omega_s < \Delta_i <  \hbar\omega_0$, removes the tendency to the 2DEHP overcooling, so that the effective temperature decreases  smoothly. This because when
the effective temperature $T$ becames sufficiently low due to the cooling effect of the high energy optical phonons, further decrease in this temperature is blocked by the energy absorption from the low energy
optical phonons (i.e., the surface optical phonons). Although their number $N_s = [\exp(\hbar\omega_s/T_0)  - 1]^{-1}\simeq \exp(-\hbar\omega_s/T_0)$ is small, it, nevertheless, exceeds the number of the GL optical phonons $N_0 = [\exp(\hbar\omega_0/T_0)  - 1]^{-1} \simeq  \exp(-\hbar\omega_0/T_0)$.

Figure~3 shows the same dependences as in Fig.~2 but 
calculated numerically for more general situations when both the GL optical phonons ($\hbar\omega_0 = 200$~meV) and the surface optical phonons ($\hbar\omega_S = 100$~meV) contribute to the relaxation processes.
As seen from Fig.~3, at the moderate injection current densities ($j \lesssim j_G$) assumed in the calculations for Fig.~2,
the carrier interaction with the surface optical phonons weakly affects the $T$ versus $j/j_G$
and $(\mu_e + \mu_h)$ versus $j/j_G$ relations at least at $s \leq 0.1$.

However, as demonstrated in Fig.~3, when $\Delta_i < \hbar\omega_0$ but $\Delta_i > \hbar\omega_S$, at larger $j/j_G$,  the surface plasmons effectively weaken the 2DEHP cooling even at relatively small strength of the carrier interaction with these plasmons (at small values of parameter $s$). When $s=1$, the effective temperature $T$ is close to $T_0$ even at rather high injection current densities. This is attributed to approximately equal contributions of the GL optical phonons to the cooling and the surface plasmons to the heating ($\hbar\omega - \Delta_i \simeq \Delta_i - \hbar\omega_S$). It is worth noting that at $\Delta_i < \hbar\omega_0$ but $\Delta_i > \hbar\omega_S$,  the carrier interaction
with the surface optical phonons does not prevent the 2DEHP degeneration and, hence, does not prevent the population inversion.

\section{DC current-voltage characteristics.}

Disregarding   the nonuniformity of the potential along the GL in the $x$-direction, 
(i. e., disregarding the current-crowding considered below in Sec.~VIII) , the device current-voltage characteristic can be found deriving $V$ as a function
of the applied voltage $U$ (see Fig.~1(b)).
Due to a smallness of the factor $D$, one can find from Eq.~(6) that in reality  $(\mu_e - \mu_h) \ll (\mu_e + \mu_h)$. 
Hence $\mu_e \simeq (\mu_e + \mu_h)/2$. Considering, in particular, the case $s \ll 1$ in which   Eqs.~(3) and (4) are valid, we find 

\begin{eqnarray}\label{eq11}
\mu_e \simeq 
\frac{\biggl(\displaystyle\frac{T_0}{2}\biggr)}{1 - \displaystyle\frac{T_0}{\hbar\omega_0}\ln\biggl[1 +\displaystyle\frac{j}{j_G}\biggl(\frac{\Delta_i}{\hbar\omega_0} - 1\biggr)\frac{1}{a}\biggr]}\nonumber\\
\times\ln\biggl[\frac{1 +  \displaystyle\displaystyle\displaystyle\frac{j}{j_G}}{1  +\displaystyle\frac{j}{j_G}\biggl(\frac{\Delta_i}{\hbar\omega_0} - 1\biggr)\frac{1}{a}}\biggr].
\end{eqnarray}

Considering Eq.~(11), one can present the current-voltage characteristic $U$ versus $j/j_G$ in the following (inexplicit) form:

\begin{eqnarray}\label{eq12}
U - \frac{\Delta_V}{e} \simeq V_0\frac{j}{j_G} +
\frac{\biggl(\displaystyle\frac{T_0}{2e}\biggr)}{1 - \displaystyle\frac{T_0}{\hbar\omega_0}\ln\biggl[1 +\displaystyle\frac{j}{j_G}\biggl(\frac{\Delta_i}{\hbar\omega_0} - 1\biggr)\frac{1}{a}\biggr]}\nonumber\\
\times\ln\biggl[\frac{1 +  \displaystyle\displaystyle\displaystyle\frac{j}{j_G}}{1  +\displaystyle\frac{j}{j_G}\biggl(\frac{\Delta_i}{\hbar\omega_0} - 1\biggr)\frac{1}{a}}\biggr].
\end{eqnarray}
Here $V_0 = d\Sigma_0/N_ab_P\tau_{Opt}^{inter}$. 
For the parameters used in above estimate, $V_0 \simeq 40$~mV.

When $\Delta_i  = \Delta_V + 3T_0/2< \hbar\omega_0$, Eq.~(12) describes a monotonically rising current-voltage characteristics tending to the saturation ($j \simeq  j_{\infty}$) at very high voltages.

If $\Delta_i < \hbar\omega_0$,  Eq.~(12) yields the following expression for the voltage corresponding to the current saturation:

\begin{eqnarray}\label{eq13}
U_{sat} = \frac{\Delta_V + \hbar\omega_0}{2e} + \frac{V_0a}{[(\Delta_V + 3T_0/2)/\hbar\omega_0 - 1)]}.
\end{eqnarray}

When the effect of the surface optical phonons is tangible, the current-voltage characteristics becomes a sublinear.

\section{Dynamic conductivity}

\begin{figure}[t]
\centering
\includegraphics[width=6.0cm]{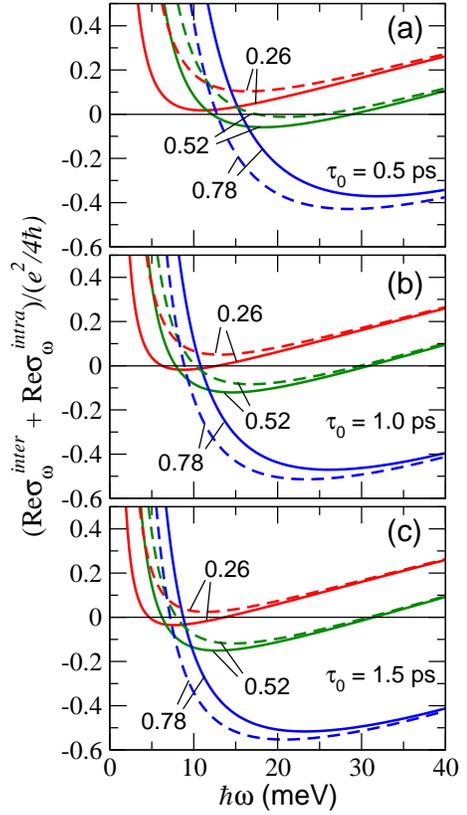}
\caption{Real part of the GL dynamic conductivity (Re$\sigma_{\omega}^{inter}$ + Re$\sigma_{\omega}^{intra}$) as a function of the radiation energy $\hbar\omega$ for different values of normalized injection current density $j/j_G$ (1 -  $j/j_G = 0.26$; \, 2 - $j/j_G = 0.52$;\,  $j/j_G = 0.78$)  and different carrier momentum relaxation times
 in the GL: (a)  $\tau_0 = 0.5$~ps
(b) $\tau_0 = 1$~ps,
and (c) $\tau_0 = 1.5$~ps ($\Delta_V = 100$~meV)  in the absence of surface optical phonon scattering, i.e.,$s = 0$ (solid lines - $\tau_p \propto \tau_0p^{-1}$ and dashed lines - $\tau_p = tau_0$).}
\label{F4}
\end{figure}

The contributions of the direct interband optical transitions and the intraband radiative transitions assisted with
the carrier scattering (leading to the Drude absorption) to the pertinent components of the GL conductivity 
$\sigma_{\omega}^{inter} = {\rm Re} \sigma_{\omega}^{inter} + {\rm Im} \sigma_{\omega}^{inter}$
and $\sigma_{\omega}^{intra} = {\rm Re} \sigma_{\omega}^{intra} + {\rm Im} \sigma_{\omega}^{intra}$ constitute the GL net dynamic conductivity. In particular,  Re$\sigma_{\omega}^{inter}$ can be found as in references
~\cite{12,57,58}: 

\begin{eqnarray}\label{eq14}
{\rm Re} \sigma_{\omega}^{inter} 
=\frac{\displaystyle \biggl(\frac{e^2}{4\hbar}\biggr)\sinh\biggl[\frac{\hbar\omega - (\mu_e + \mu_h)}{2T}\biggl]}
{\displaystyle\cosh\biggl[\frac{\hbar\omega - (\mu_e + \mu_h)}{2T}\biggl] + \displaystyle\cosh\biggl(\frac{\mu_e - \mu_h}{2T}\biggl)}\nonumber\\
\simeq \frac{\displaystyle \biggl(\frac{e^2}{4\hbar}\biggr)\sinh\biggl[\frac{\hbar\omega - (\mu_e + \mu_h)}{2T}\biggl]}
{\displaystyle\cosh\biggl[\frac{\hbar\omega - (\mu_e + \mu_h)}{2T}\biggl] + \displaystyle\cosh\biggl[\frac{T_0^2D}{2T(\mu_e + \mu_h)}\frac{j}{j_G}\biggl]},
\end{eqnarray}
Up to fairly large values of $j/j_G$, the argument of the first  $\cosh$-function in the denominator of the expression in the right-hand side of Eq.~(14)  is much larger than that in the second $\cosh$-function. Taking this into account, Eq.~(14) can be reduced to the standard form~\cite{12}:

\begin{equation}\label{eq15}
{\rm Re} \sigma_{\omega}^{inter} \simeq \frac{e^2}{4\hbar}\tanh\biggl(\frac{\hbar\omega - \mu_e - \mu_h}{4T}\biggl).
\end{equation}
The quantity Im$\sigma_{\omega}^{inter}$ can be presented as~\cite{57}

\begin{eqnarray}\label{eq16}
{\rm Im} 
\sigma_{\omega}^{inter} 
=i \biggl(\frac{e^2}{4\hbar}\biggr)
\frac{4\hbar\omega}{\pi}
\int_0^{\infty} \frac{G(\varepsilon) - G(\hbar\omega/2)}{(\hbar\omega)^2 - 4\varepsilon^2}
d\varepsilon,
\end{eqnarray}
where $G(\varepsilon) = \tanh[2\varepsilon - (\mu_e + \mu_h)/4T]$.

The intraband contributions  Re~$\sigma_{\omega}^{intra}$ + Im~$\sigma_{\omega}^{intra}$
depend on the carrier momentum relaxation  mechanisms in the GL, particularly, on the range of the effective
carrier-carrier interactions  and on disorder~\cite{59} (see also~\cite{50}).
 At fairly high  carrier densities,
 expected  under the injection conditions under consideration, 
the electron-hole  interactions are the main mechanism of the  momentum relaxation~\cite{60,61,62}.
Due to special features of the mutual scattering of the carriers with the linear dispersion law~\cite{59,60,61},
 such  scattering is a short range scattering. The mutual carrier scattering is similar to the scattering on uncharged and screened charged impurities, as well as the acoustic phonons and defects. In this case, the momentum relaxation time
as a function of the electron or hole momenta can be presented as $\tau_p = \tau_0(p_0/p)$~\cite{50,51}, where $p_0 =T_0/v_W$
and $\tau_0$ is the characteristic carrier  momentum relaxation time. If the dominant scattering mechanism is associated with the carrier interactions with weakly screened charged impurities or their clusters, i.e.,  with the long-range scatterers, one can set $\tau_p = \tau_0(p/p_0)$.
When the interaction with  both the short- and long-range scatterers is important,  the approximation
$\tau_p = \tau_0 = const$ could be used~\cite{12,17,57,63}.
Considering this, 
one can arrive at

\begin{eqnarray}\label{eq17}
{\rm Re} \sigma_{\omega}^{intra} + {\rm Im} \sigma_{\omega}^{intra} 
= \biggl(\frac{e^2}{4\hbar}\biggr)
\frac{8\langle \varepsilon_p\rangle\tau_0}{\pi\hbar(1 - i\omega \langle \tau_p\rangle)},
\end{eqnarray}
where  $\langle \varepsilon_p \rangle = T_0$, $\langle \tau_p \rangle = (2\tau_0T_0)/(\mu_e + \mu_h)$  at $\tau_p \propto p^{-1}$  and 
$\langle \varepsilon_p \rangle = (\mu_e + \mu_h)/2$, 
 $\langle \tau_p \rangle = \tau_0$ at $\tau_p = \tau_0 = const$  (valid when $\mu_e + \mu_h > T$) .

At $\hbar\omega < \mu_e + \mu_h$, Eqs.~(14) and (15) yields Re$~\sigma_{\omega}^{inter} < 0 $.

If the dominant  scattering mechanism of the electrons and holes in the GL is their mutual interaction,
the  quantity $\tau_0$ calculated for $T_0 = 25$~meV and $\kappa = 6$ (for a GL sandwiched between the PL and hBN) is about of $\tau_0 = 3.6$~ps~\cite{62}.
Accounting for other scattering mechanisms (impurities, acoustic phonons, and so on), one can set $\tau_0 = 1$~ps.
Assuming $1.0 - 3.6$~ps,  the net real part of the dynamic conductivity is negative
in the frequency range $\omega/2\pi \geq (3.44 - 6.50)$~THz.

Figure~4 shows the spectral dependences of the real part of the net dynamic conductivity in the GL
 (Re ~$\sigma_{\omega}^{inter}$ + Re~$\sigma_{\omega}^{intra}$) calculated for  the cases $\tau_p \propto \tau_0p^{-1}$ (solid lines) and $\tau_p = tau_0 = const$ (dashed lines) using Eqs.~(15) and (17)
with Eqs.~(3) and (4) for $T$ and $(\mu_e + \mu_h)/T$ for different characteristic momentum relaxation $\tau_0$ and different values of the normalized injection current density $j/j_G$. Other parameters used are $T_0 =  300$~K, $\hbar\omega_0 = 200$~meV, $\Delta_V = 100$~meV, 
(for $\kappa \simeq 6$), $a = 0.25$,  and $s \ll 1$.

As seen from Fig.~4, the real part of the dynamic conductivity of the 2DEHP can be negative at sufficiently strong injection pumping in a certain range of  $\hbar\omega$ (compare the curves for $j/j_G = 0.52$ and $j/j_G = 0.78$.  An increase in the injection current density leads to  the reinforcement of the negative 
dynamic conductivity and widening of the range where this conductivity is negative.
This is mainly due to the rise of  Re$\sigma_{\omega}^{inter}$ when the net quasi-Fermi energy
$(\mu_e + \mu_h)$ increases [see Eq.~(15)].
The comparison of the solid and dashed lines (corresponding to different momentum dependences of the momentum
relaxation time) shows that 
they are rather close, although the character of the carrier scattering plays some role.
The fact that the 
hBN substrate is virtually free of charged impurities (providing the long-range carrier scattering), 
is in favor of  the dependence $\tau_p \propto \tau_0p^{-1}$. Therefore, calculating plots in  the  consequent figures, we set $\tau \propto \tau_0p^{-1}$.

Figure~5 shows the spectral dependences of the real part of the 2DEHP dynamic conductivity similar
to those in Fig.~4, but obtained for a higher value of the surface optical phonon parameter $s$, namely for $s = 0.1$. Comparing the plots of Figs.~4 and 5, one can see that an increase in the parameter $s$ results in a weakening of the negative dynamic conductivity effect.
Enhancing the carrier mobility in the GL, i.e., and increase in $\tau_0$ can markedly reinforce the negative dynamic conductivity, due to weakening of the intraband absorption. As follows from Fig.~3(c),
the quantity $\mu_e + \mu_h)/T$ can markedly exeed unity even $s \sim 1$, but at relatively high injection current densities  ($j/j_G \sim 3 -4$). This implies that the effect of the negative dynamic
conductivity can pronounced in the case of relatively strong carrier interaction with the surface optical phonons as well.

\begin{figure}[t]
\centering
\includegraphics[width=6.0cm]{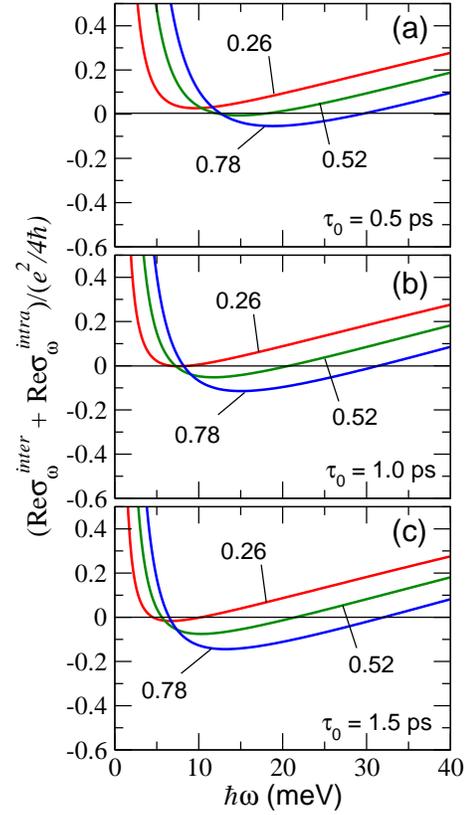}
\caption{The same as in Fig. 4 but for surface optical phonons parameter $s = 0.1$. }
\label{F5}
\end{figure}

\section{Surface plasmons amplification coefficient}

Using the equations for the GL dynamic conductivity under the injection pumping given in the Sec.~VI, invoking the Maxwell equations, 
considering the structure geometry, and following the method applied previously~\cite{17,18,22}, one can derive  the  dispersion equation for the surface plasmons with the frequency $\omega$, in which the ac electric and magnetic fields components are proportional to $\displaystyle \exp\biggl(i\rho\frac{\omega}{c}y -i\omega\,t\biggr)$ 
propagating in the direction parallel to the side contacts (along the axis $y$). Assuming (see Sec. VIII) that the plasmon absorption in the PL is due to the interaction with the  holes (Drude absorption), one can arrive to the following dispersion equation:

\begin{eqnarray}\label{eq18}
\varepsilon_{hBN}\sqrt{\varepsilon_z - \rho^2} + \varepsilon_z\sqrt{\varepsilon_{hBN} - \rho^2}\nonumber\\
 + \frac{4\pi}{c}\sigma_{\omega}\sqrt{\varepsilon_z - \rho^2}\sqrt{\varepsilon_{hBN} - \rho^2} = 0
\end{eqnarray}
with

\begin{equation}\label{eq19}
\varepsilon_z = \varepsilon_P\biggl(1  -\frac{\omega_P^2}{\omega^2  + i \gamma_P\omega}\biggl). 
\end{equation}
Here $\sigma_{\omega} = \sigma_{\omega}^{inter} +
\sigma_{\omega}^{intra}$ is the GL net dynamic conductivity, 
the  low-frequency dielectric constants of  the hBN
$\varepsilon_{hBN}$ is  taken from~\cite{64,65},
 $\omega_P = \sqrt{4\pi\,e^2N_a/m\varepsilon_P}$ is the plasma frequency of holes in the PL, $\gamma_P =e/mb_P$ is the plasma oscillation damping constant associated with the Drude absorption in the PL, and $c$ is the speed of light in vacuum.
The quantities Re$(\rho)$ and $2\omega$Im$(\rho)/c$, obtained from the solution of Eq.~(18), are the plasmon propagation index and the plasmon absorption or amplification coefficient (depending on the sign) , respectively. 
Deriving the dispersion equation for the surface plasmons, we have accounted for
the interaction of the electromagnetic radiation with phonons in PLs resulting in the single-phonon absorption if and only if the radiation is polarized along the axis $z$. The pertinent absorption coefficient is two order of magnitude smaller than that in the standard polar semiconductors, although there is a narrow peak at 14~THz with the absorption coefficient about 500~cm$^{-1}$. The two-phonon absorption is relatively week (about 15~cm$^{-1}$ in the range 7.5 - 14 THz~\cite{66}).
Therefore, the Drude mechanism plays the main role in the plasmon absorption in the PL as was assumed above.

Figure~6 shows the spectral dependences   of the plasmon amplification coefficient $ \alpha_{\omega} = - 2\omega {\rm Im}\rho/c$.
We assumed that the acceptor density in the BL and the thickness of this layer
are  equal to $N_a = 5\times 10^{15}$~cm$^{-3}$
and  $d = 10^{-4}$cm, respectively. The injection current densities and 
other ther parameters are the same as for Fig.~5.
As seen from Fig.~5, in the frequency  range where the 2DEHP dynamic conductivity is negative,
the amplification coefficient can be fairly large,  of the order of $\alpha_{\omega} \simeq (1.5 - 2.0)\times 10^{4}$~cm$^{-1}$. The large amplification coefficient of the plasmonic mode  in comparison with the photonic modes is attributed to a small plasmon propagation velocity compared to the speed of light. 

As seen from Fig.~3, the reinforcement of the surface optical phonon scattering (increase in $s$)  gives rise to  pronounced variations of $T$ and $(\mu_e + \mu_h)$ and, hence, $\alpha_{\omega}$.
Figure~7 shows the $\alpha_{\omega}$ versus $\hbar\omega$ calculated for different $s$.
An increase in $s$ corresponds to a drop of $\alpha_{\omega}$. As seen, at  $j/j_{G} = 0.78$ and $s \geq 0.60$, $\alpha_{\omega}$ becomes negative. However, for a larger  $j/j_G$,  $\alpha_{\omega}$
can be positive at a larger $s$.

The obtained values of the amplification coefficient
are close to those in the GL-based structures with the side double injection. This is because the Drude absorption in the BL is relatively weak, at least, at $N_a \leq 5\times 10^{15}$~cm$^{-3}$. 
At a higher doping of the PL, this absorption can decrease $\alpha_{\omega}$ even leading
to the transition from the amplification to the damping of the plasmonic modes as shown in Fig.~9.
A  weak Drude absorption is partially associated with strong localization of the y-and z-components of the plasmon electric field around the GL. The latter is demonstrated in 
Fig.~7. A strong localization of the plasmon electric field far from the contact p$^+$-PL (at the distance about $1~\mu$m) prevents  the plasmon damping due to the absorption in this layer.

\begin{figure}[t]
\centering
\begin{scriptsize}
\end{scriptsize}\includegraphics[width=6.0cm]{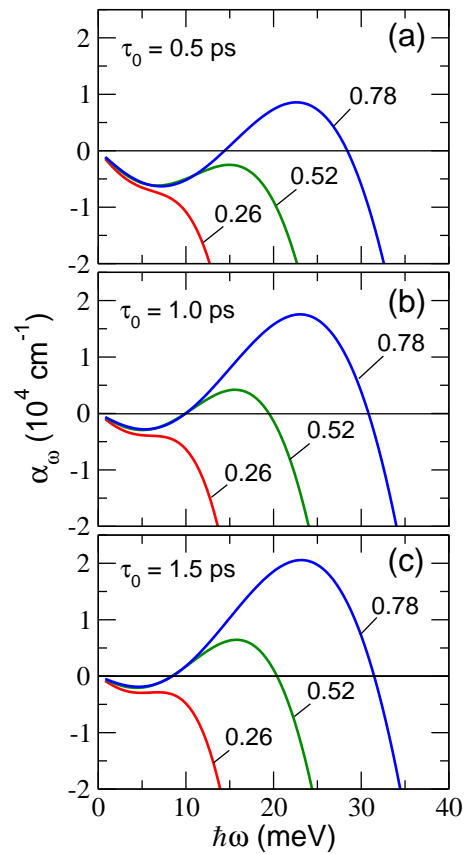}
\caption{Spectral characteristics of the plasmon amplification coefficient $\alpha _{\omega} = - 2\omega {\rm Im}\rho/c$ at $j/j_G = 0.26,\,  052,$  and 078: $N_a = 5\times 10^{15}$~cm$^{-3}$, $d = 10^{-4}$~cm, other parameters are the same as for. Fig.~5.
}
\label{F6}
\end{figure}

\begin{figure}[t]
\centering
\includegraphics[width=6.0cm]{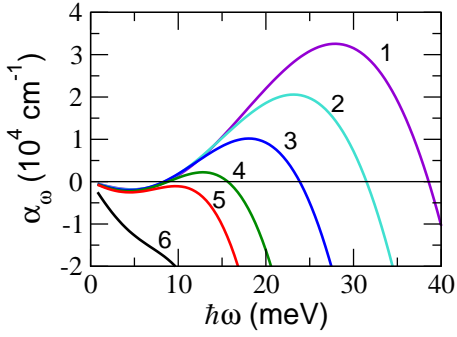}
\caption{Spectral dependence of the absorption coefficient for different values of parameter $s$
(1 - $s=0.05$,\, 2 - $s =0.10$,\, 3 - $s= 0.20$,\, 4 - $s=0.40$,\, 5 - $s=0.6$,\ 6 - $s=0.80$) 
and $\Delta_V = 100$~meV, $\tau_0 = 1.5$~ps, $j/j_G = 0.78$,  and $N_a = 5\times 10^{15}$~cm$^{-3}$.
}
\label{F7}
\end{figure}

\begin{figure}[t]
\centering
\includegraphics[width=6.0cm]{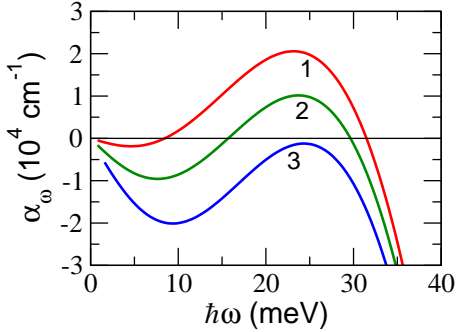}
\caption{Spectral dependence of the absorption coefficient for different acceptor densities $N_a$ 
(1 - $N_a = 5\times10^{15}$~cm$^{-3}$,\, 2 - $N_a = 5\times10^{16}$~cm$^{-3}$,\,and 
3 - $N_a = 1\times 10^{17}$~cm$^{-3}$) 
in the PL injector: $s = 0.1$ and the same parameters as for Fig.~7.
}
\label{F8}
\end{figure}

\begin{figure}[h]
\centering
\includegraphics[width=6.0cm]{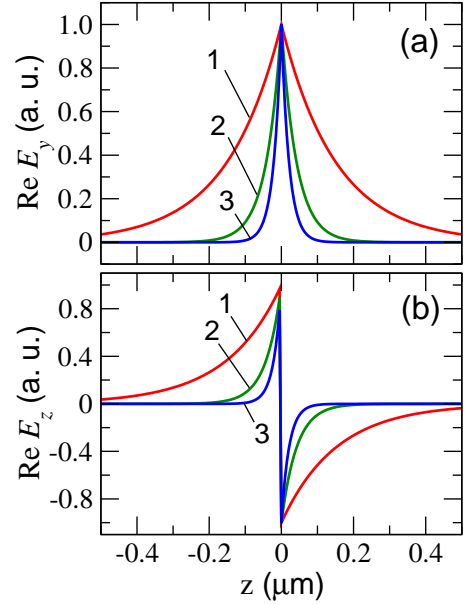}
\caption{Spatial distributions along the $z$-direction perpendicular to the GL and BL plane
of the plasmonic mode electric field components for different plasmon energies (1 - $\hbar\omega = 10$~meV,\,
 2 - $\hbar\omega = 20$~meV,\, and  1 - $\hbar\omega = 30$~meV):
$s = 0.1$ and the same other parameters as for Fig.~7.
}
\label{F9}
\end{figure}

 \section{Discussion}
\subsection{Role of the Auger processes}


The interband Auger processes decrease the split of the electron and hole quasi-Fermi energies
$(\mu_e+\mu_h)$.
At low injection current densities $j \ll j_G$, the rate of the Auger recombination can be taken to be proportional to $(\mu_e+\mu_h)/T_0\tau_A$. 
The variation of this energy associated with the Auger processes can be estimated as $\varepsilon_{Auger} \sim T, \mu_e, \mu_h \ll \hbar\omega_0$, hence 
 the contribution of the Auger processes to the 2DEHP energy balance can be disregarded. Considering this and using the linearized Eqs.~(1) and (2), we arrive at 
 
\begin{equation}\label{eq20}
\frac{\mu_e + \mu_h}{T} + (1 + a)\hbar\omega_0\biggl(\frac{1}{T_0} - \frac{1}{T}\biggr)
 = 
\frac{j}{j_G}\frac{\Delta_i}{\hbar\omega_0}
\end{equation}
The equation governing the electron and hole  balance  is given by:

\begin{equation}\label{eq21}
\frac{\mu_e + \mu_h}{T} + \frac{\hbar\omega_0}{(1 + a_A)}\biggl(\frac{1}{T_0} - \frac{1}{T}\biggr)
 = \frac{j}{j_G}\frac{1}{(1 + a_A)}.
\end{equation}
where
$a_{A} = \tau_{Opt}^{inter}/\tau_{A}$ can be called as  the Auger parameter, which can be estimated using~\cite{56} (see also references therein).
 Equations~(20) and (21) result in
 
 \begin{equation}\label{eq22}
\frac{T-T_0}{T_0} \simeq  \biggl[\frac{\Delta_i(1+a_A)/\hbar\omega_0 - 1}{a + a_A + aa_A}\biggr]\frac{j}{j_G},
\end{equation}

\begin{equation}\label{eq23}
\frac{\mu_e + \mu_h}{T_0} \simeq \biggl(\frac{1 + a - \Delta_i/\hbar\omega_0}
{a + a_A + aa_A}\biggr) \frac{j}{j_G}.
\end{equation}

 At relatively weak Auger processes ($a_A \ll 1$), Eqs.~(22) and (23)  lead to the same
 dependences $(T - T_0)$ and $(\mu_e + \mu_h)$ on the injection current density $j$ as obtained
 in Sec.~III  (for the relaxation on the GL optical phonons at small $j/j_G$).
 
 Generally speaking, Eqs.~(22) and (23) show that the  Auger processes
result in slowing down the cooling (which can occur at $\Delta_i < \hbar\omega_0$) of the 2DEHP with increasing injection current.

If the Auger parameter   $a_A$ is sufficiently large ($a_A = \hbar\omega_0/\Delta_i -1$),
the cooling gives way to the heating.  At both cooling and heating og the 2DEHP,
the splitting of the quasi-Fermi energies, i.e., the quantity $(\mu_e + \mu_h)$, increase when $j$ increases providing that $\Delta_i/\hbar\omega_0 < 1 +a$.

\subsection{Heating of optical phonons}

The recombination and the intraband energy relaxation lead to the generation of nonequilibrium
(hot) optical phonons
The generated hot optical phonons cool down through anharmonic decay to acoustic phonons which are subsequently absorbed into the substrate~\cite{54,66,67,68}.
Direct cooling of the charge carriers  also  occurs via emission of  the surface phonons of the underlying polar substrate.

As demonstrated experimentally ,  the optical phonon decay time in the GL-hBN heterostructures is about~\cite{54} $\tau_{Opt}^{decay} \sim 0.200 -- 0.375$~ps, i. e., is relatively short.
At such short decay times, the deviation of the optical phonon system from equilibrium is insignificant, i.e., this system temperature $T_{Opt} \simeq T_0$.
This justifies the omission of this effect in the model used above.
 An example of the inclusion of the optical phonon heating into a similar model could be found  in~\cite{16,32}.  Due to the large specific heat capacity of hBN, the rise of the lattice temperature even under relatively strong pumping is small ($\sim 1$~K)~\cite{54}.

\subsection{Current crowding in the GL}

The finiteness of the GL conductivity can lead to a nonuniformity of the potential distribution $\varphi = \varphi(x)$ along the conductivity plane and,
consequently, to  a nonuniformity of the injection current $j = j(x)$, where axis $x$ is
in the  direction connecting the   n$^+$-contacts (see Fig.~1). This effect is akin to the current-crowding effect in the bipolar transistors and light-emitting diodes, dominating at  high current densities~\cite{69,70}.
The current crowding slows down  the $j$ versus $U$ dependence. 
The general consideration of the  current crowding  requires a rather complex mathematical modelwith  nonlinear differential equations describing  the   potential  and current density distributions. This is beyond the scope of the present paper. Here we limit ourselves to the case when the current crowding is not too strong
and find the pertinent conditions.

Since the resistance of the side contacts to the GL appears to be not a challenging issue~\cite{71,72,73,74},
we disregard the contribution of the contact  resistance to the net potential drop, $U$, between the p$^+$-contact and the n$+$-side contacts.
The lateral variation of the injection current density in the in-plane direction $x$ (see Fig.~1)  can  be approximately found from the continuity equation:

\begin{equation}\label{25}
\frac{d^2j}{dx^2} = K^2j
\end{equation}
with the boundary conditions $j = j_0|_{x = \pm l}$ given at the side contact edges ($x = \pm l$).
Here  $2l$ is the spacing between the side-contacts to the GL, $j_0$ is given by Eq.~(11), and
 $K \simeq  \sqrt{(b_B/b_G)(N_a/\Sigma_Gd)}$,
$b_G$ and $\Sigma_G$ are the mobility and density of the carriers  in the GL, respectively.

Solving Eq.~(26), we find

 \begin{equation}\label{25}
j = j_0\frac{\cosh(Kx)}{\cosh(Kl)}.
\end{equation}

The value of the injection current density sag $\delta j =[1 - \cosh^{-1}(Kl)] \simeq (Kl/2)^2$ is relatively small if   $2l \ll L = 4K^{-1} = 4\sqrt{(b_G/b_P)(\Sigma_Gd/N_a)}$. This inequality implies that the lateral
resistance of the GL is much smaller that the vertical resistance of the PL.
 Assuming $N_a = 5\times 10^{15}$~cm$^{-3}$, $\Sigma_G = 10^{12}$~cm$^{-2}$, $d = 10^{-4}$~cm,$b_G = 10,000$~cm$^2$/V$\cdot$s, for $b_P = (250-500)$~cm$^2$/V$\cdot$, we obtain  that
the current density nonuniformity can be disregarded if  $2l \ll L = (25 - 36)\times \mu$m. 
Larger values of $2l$ correspond to the smaller contact leakage currents~\cite{30}.
The latter inequality corresponds to the real device sizes.

On the contrary, in the GL- heterostructures with the lateral electron and hole double  injection from the  side contacts~\cite{30}, the lateral nonuniformity of the carrier densities is determined by the diffusion length $L_D$. The latter is about a few micrometers. 
Since  $L \gg L_D$,  the GL-PL heterostructures with the combined injection can provide the negative dynamic conductivity 
in much larger area than the heterostructures with the lateral injection. This implies that the THz sources based on the GL-PL heterostructures  can demonstrate markedly higher output power.

\section*{Conclusion}
We proposed the p$^+$PL-PL-GL heterostructures with the lateral electron and vertical hole injection as the the active elements of the plasmonic lasers. Using the developed device model, we calculated the effective temperature of the carriers, their quasi-Fermi energies, and the dynamic conductivity
of the 2DEHP in the GL.  Under sufficiently strong injection current densities, the dynamic conductivity can be negative in a certain range of the plasmon energies  providing positive and a fairly large amplification coefficient of the plasmonic mode. Due to a relatively small energy of the holes injected from
the PL injecting contact  in comparison with the optical phonon energy in the GL, the carrier effective temperature can be lower than the ambient temperature. This, together with the possibility  of the negative dynamic conductivity realization in fairly large GL areas,  promotes a more efficient THz lasing.
Similar GL-based heterostructures can  include the black arsenic injecting layers and other injecting layer materials with a proper band alignment to the GLs~\cite{75,76}. Using the substrates providing weaker energy
and momentum carrier relaxation in the GL (instead of hBN considered above, one can achieve a stronger negative dynamic conductivity and higher  amplification amplification of the plasmonic modes at a weaker  injection.
The plasmonic lasing can be enabled by the plasmon reflection from the end faces and by the realization
of the distributed feedback using the highly conducting saw-tooth (serrated) side contacts~\cite{26}.

The author are grateful to V. Leiman, V. Mitin, A. Arsenin, and S. V. Morozov for fruitful and stimulating 
discussions.
The work at RIEC and UoA
was supported by Japan Society for Promotion of Science (Grants Nos. 16H-06361 and
16K14243). The joint work at RIEC and IPM and the work at
 RPI were 
supported by Russian Foundation for Basic Research (Grant No. 18-52-50024) and 
by
Office of Naval Research (Project Monitor
Dr. Paul Maki), respectively.

\section*{Appendix A. Injection current into the GL}
\setcounter{equation}{0}
\renewcommand{\theequation} {A\arabic{equation}}

The injected current  coincides with the current across the p-PL 
in the hole injector.  At low bias voltages, the injected current is associated with the hole diffusion across
the  BL.  When $V = V_{bi} \simeq \Delta_V$, i.e., when $U = 0$ , its density can be estimated as $j_0 \simeq e D_PN_a/d = eb_PT_0N_a/d$. Here $D_P$ and $b_P$ are  the hole diffusion coefficient and mobility in the PL 
perpendicular to its plane (perpendicular to the atomic sheets forming
the PL layers structure) and $N_a$ the acceptor density in this layer.

 At larger values of $|U|$, when the  voltage drop
across the PL $V > |U|- V_{bi} - \mu_e/e > 0$ [see Fig.~1(b)], i.e., in the operation regime,
the injected  current is determined by the PL resistance.
Taking into account that the holes in the p-PL should not be heated too strongly,
we assume that the average electric field in this layer $E = V/d$ is moderate, where $d$ is the thickness of the PL. The acceptor density in the PL  can be set $N_a \sim (2 - 5)\times 10^{15}$~cm$^{-3}$~\cite{34,35}.
In such a situation, the hole density in the PL at moderate voltages $p \simeq N_a$, and the current density across the PL $J$ (which coincides with the density of the  recombination current in the GL) is given by

\begin{equation}\label{A1}
j= \frac{V}{d\rho_P}, \qquad \rho_P = \frac{1}{eN_ab_P}.
\end{equation}
Here  $\rho$ is the PL resistivity.

Setting the acceptor density in the PL $N_a \sim (2 - 5)\times 10^{15}$~cm$^{-3}$~\cite{34,35},  $b_B = (250 -500)$~cm$^2$/V$\cdot$s, $d = 10^{-4}$~cm, 
we obtain $j_0 \simeq 8 - 20$~A/cm$^2$. If
$V = (0.1 - 1.0)$~V,
we obtain $j = 2\times10^2 - 4\times10^3)$~A/cm$^2$. Since at the normal device operation $j_0 \ll j$,
we can neglect $j_0$

The hole effective temperature in the PL $T_B$ can be estimated using the following equation:

\begin{equation}\label{A2}
N_a\frac{(T_P - T_0)}{\tau_P^{\varepsilon}} = j\frac{V}{d},
\end{equation}
so that 
  
\begin{equation}\label{A3}
T_P =T_0 + \frac{\tau_P^{\varepsilon}}{eb_PN_a^2}j^2 = T_0 + \frac{m}{e^2N_a^2}\frac{\tau_P^{\varepsilon}}{\tau_P^{p}}\,j^2.
\end{equation}
Here $\tau_P^{\varepsilon}$ and $\tau_P^{p}$ are the hole energy and momentum relaxation times in the PL. 
Considering Eq.~(A3), one can find that 

\begin{equation}\label{A4}
\Delta_i = \Delta_V + \frac{3T_0}{2}\biggl[1 + \Theta\biggl(\frac{j}{j_G}\biggr)^2\biggr],
\end{equation}
where 
$$
\Theta = \frac{m}{N_a^2T_0}\frac{\tau_P^{\varepsilon}}{\tau_P^{p}}\biggl(\frac{\Sigma_0}{\tau_{Opt}^{inter}}\biggr)^2.
$$

Deriving the hole momentum relaxation time $\tau_P^{p}$ from the value of the hole mobility $bP$ 
($\tau_P^{p} \simeq (0.4 - 0.8)\times 10^{-13}$~s) with $m= 2.5\times10^{-28}$,  setting $\tau_P^{\varepsilon} \simeq 10\tau_P^{p}$  and $\Sigma_0/\tau_{Opt}^{inter} = 10^{21}$~cm$^{-2}$s$^{-1}$, 
for $N_a = 5\times 10^{15}$~cm$^{-3}$, one obtains   $\Theta \simeq 2.4\times 10^{-3}$.
The latter estimate implies that in the range of realistic current densities one can put $\Delta_i = \Delta_V +3T_0/2 \simeq 137$~meV.

\section*{Appendix B. Nondegenerate electron-hole system}
\setcounter{equation}{0}
\renewcommand{\theequation} {D\arabic{equation}}

When $|\mu_e|, |\mu_h| < T$, the electron-hole system in the GL is non-degenerate, so that

\begin{equation}\label{eqB1}
\Sigma_e \simeq \frac{2T^2}{\pi\hbar^2v_W^2}\exp\bigg(\frac{\mu_e}{T}\biggr), \qquad
\Sigma_h \simeq \frac{2T^2}{\pi\hbar^2v_W^2}\exp\bigg(\frac{\mu_h}{T}\biggr).
\end{equation}
As a result, taking into account Eq.(8), instead of Eq.~(9) we obtain  

\begin{equation}\label{eq22}
\mu_e - \mu_h \simeq T_0\frac{D}{2} \frac{j}{j_G}
\end{equation}

\begin{equation}\label{eq23}
\mu_e + \mu_h \simeq T_0\biggl[1 - \biggl(\frac{\Delta_i}{\hbar\omega_0} - 1\biggr)\frac{1}{a}\biggr] \frac{j}{j_G},
\end{equation}

At $V = 0.1$~V, $N_a = 5\times10^{15}$~cm$^{-3}$,  $d = 1.0~\mu$m, $\kappa = 6$ 
one obtains $j \simeq 1.6\times (10^3 -  10^4)$~A/cm$^2$.
This yields, $(\mu_e + \mu_h)/T \simeq 2.3 - 4.6$ and $(\mu_e - \mu_h)/T  \ll 1 $.

\end{document}